\pdfoutput=1
\documentclass[11pt]{article}
\usepackage{latexsym,cite,amssymb,amsmath,amsfonts}
\usepackage{enumitem}
\usepackage{float}
\usepackage{xcolor}
\usepackage{setspace,hypernat,graphicx,epstopdf}

\definecolor{MyDarkBlue}{rgb}{0.15,0.15,0.45}
\usepackage[linktocpage=true]{hyperref}
\hypersetup{
colorlinks=true,
citecolor=MyDarkBlue,
linkcolor=MyDarkBlue,
urlcolor=MyDarkBlue,
pdfauthor={},
pdftitle={},
pdfsubject={hep-th}
}

\makeatletter

\@addtoreset{equation}{section} \makeatother

\usepackage{hyperref}

\newcommand{\eref}[1]{Eq.\,(\ref{#1})}

\newcommand{\half}{{{\textstyle\frac{1}{2}}}}

\newcommand{\be}{\begin{equation}}
\newcommand{\ee}{\end{equation} }
\newcommand{\beqa}{\begin{eqnarray} }
\newcommand{\eeqa}{\end{eqnarray} }
\newcommand{\ba}{\begin{array}}
\newcommand{\ea}{\end{array}}

\newcommand\tr{{\rm \, tr}}
\newcommand\Tr{{\rm Tr}}

\newcommand\cL{{\cal L}}

\begin{document}

\thispagestyle{empty}

\renewcommand{\thefootnote}{\fnsymbol{footnote}}

\rightline{CERN-PH-TH/2012-178}
\rightline{KCL-MTH-12-07}

\vspace{1.8truecm}

\vspace{15pt}

\centerline{\LARGE \bf {\sc Periodic Arrays of M2-branes}    }

 \vskip0.8cm
\centerline{ 
{\large {\bf {\sc Imtak~Jeon${}^{\,a,b,}$}}}\footnote{E-mail address: \href{mailto:imtak@sogang.ac.kr}{\tt imtak@sogang.ac.kr}}, 
{\large {\bf {\sc Neil Lambert${}^{\,b,c,d}$}}}$^,$\footnote{E-mail address: \href{mailto:neil.lambert@cern.ch}{\tt neil.lambert@cern.ch}} 
{ and} 
{\large {\bf{\sc Paul Richmond${}^{\,d,}$}}}\footnote{E-mail address: \href{mailto:paul.richmond@kcl.ac.uk}{\tt paul.richmond@kcl.ac.uk}}
}

\vspace{0.8cm}
\centerline{${}^a${\it Department of Physics, Sogang University}}
\centerline{{\it Seoul, Korea}}
\vspace{.7cm}
\centerline{${}^b${\it Theory Division, CERN}}
\centerline{{\it 1211 Geneva 23, Switzerland}}
\vspace{.7cm}
\centerline{${}^c${\it Isaac Newton Institute for Mathematical Sciences}}
\centerline{{\it 20 Clarkson Road, Cambridge CB3 0EH, UK}}
\vspace{.7cm}
\centerline{${}^d${\it Department of Mathematics, King's College London}}
\centerline{{\it The Strand, London WC2R 2LS, UK}}

\bigskip
\begin{center}
{\bf {\sc Abstract}}
\end{center}

We consider  periodic arrays of M2-branes in the ABJM model in the spirit of a circle compactification to  D2-branes in type IIA string theory.  The result is a curious formulation of three-dimensional maximally supersymmetric Yang-Mills theory in terms of fermions, seven transverse scalars, a non-dynamical gauge field and an additional scalar `dual gluon'. Upon further T-duality on a transverse torus we obtain a non-manifest-Lorentz-invariant description of five-dimensional maximally supersymmetric Yang-Mills. Here the additional scalar field  can be thought of as the components of a two-form along the torus. This action can be viewed as an M-theory description of M5-branes on ${\mathbb T}^3$.

\newpage

\renewcommand{\thefootnote}{\arabic{footnote}}
\setcounter{footnote}{0}

\setcounter{page}{1}

\section{Introduction}

One of the early results of the M2-brane theories \cite{Bagger:2006sk,Bagger:2007jr,Gustavsson:2007vu,Bagger:2007vi,Aharony:2008ug}\footnote{For a review see \cite{Bagger:2012jb}.} was that their relation to D2-branes   arises by a `novel Higgs mechanism' \cite{Mukhi:2008ux} where, far out on the Coulomb branch, the non-dynamical gauge fields `eat' a scalar so that the theory is described at low energy - low compared to the vacuum expectation value (vev) on the Coulomb branch -  by three-dimensional, maximally supersymmetric Yang-Mills (3D-MSYM). On the other hand, at least
naively, the most straightforward way to reduce from M2-branes to D2-branes is to compactify one transverse dimension on a circle. This can be done by considering an infinite array of M2-branes with equal spacing between them along some direction. Such  arrays of D-branes were considered in \cite{Taylor:1996ik} within the context of T-duality and therefore it is of interest to extend this discussion to the case of M2-branes.

At first this would seem to be a clear-cut and well-defined goal. After all the ABJM model \cite{Aharony:2008ug} allows us to consider an arbitrary number of M2-branes located in any configuration in ${\mathbb C}^4/{\mathbb Z}_k$. We can therefore use this to describe an infinite periodic array. In particular the vacuum is described by the scalar field vev:
 \be\label{config}
\langle Z^{A'}\rangle =0\ ,\qquad \langle Z^4\rangle = 2\pi i R \left(\begin{array}{ccccc} \ddots && && \\  &1&&& \\ &&0&&\\ &&&-1& \\ && && \ddots   \end{array}\right) \ ,
 \ee
where $A'=1,2,3$.  Note that each entry should be viewed as multiplying an $M\times M$ identity matrix corresponding to  $M$ M2-branes located at each site.  This configuration is illustrated in Figure 1, where we have also indicated the action of the inherent ${\mathbb Z}_k$ orbifold. 
\begin{figure}
\centering
\includegraphics[width=0.8\textwidth]{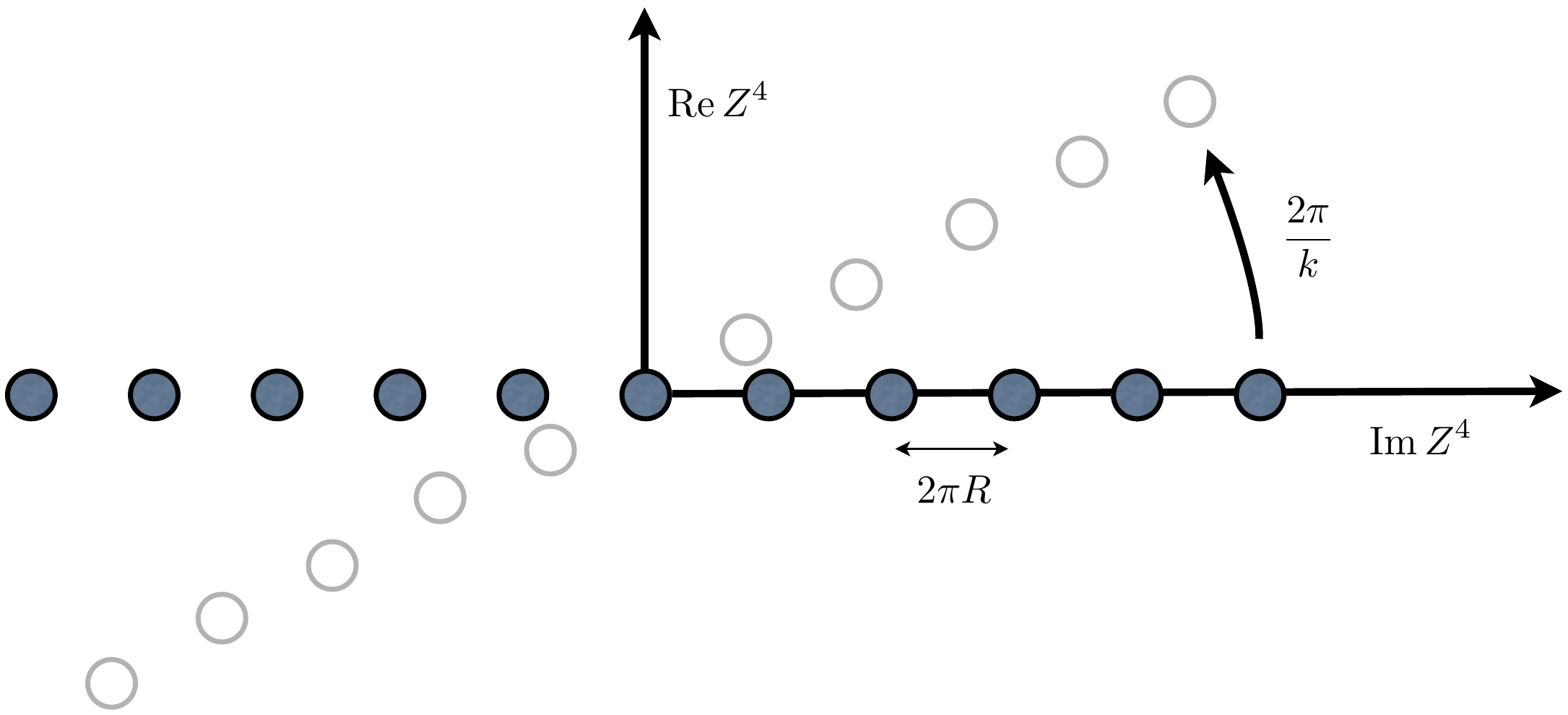}
\caption{The array of M2-branes\label{figure1}}
\end{figure}
One might worry about the effect of this orbifold  however this could in principle be avoided by taking $k=1,2$. Although this is strongly coupled we might expect to recover  a weak coupling expansion by taking the periodicity $2\pi R$ small and thus reducing to type IIA string theory.

On further reflection more serious difficulties present themselves. Although the vacuum configuration of an infinite periodic array is readily accommodated for in the ABJM model, the dynamics of the array come from considering an additional orbifold that imposes a discrete translational invariance along the array, as was done for D-branes in \cite{Taylor:1996ik}. But in the ABJM model this translational invariance is broken and, even for $k=1$, it is not a symmetry of the Lagrangian.  Rather, the restoration of this shift symmetry at $k=1$   is through non-perturbative effects involving monopole ('t Hooft) operators \cite{Bashkirov:2010kz}. Another issue is that the classical Lagrangian analysis gives spurious massless fluctuations whenever two M2-branes lie at the same distance from the origin, which are expected to be lifted by non-perturbative effects \cite{Martinec:2011nb}\footnote{Although in the case at hand this problem seems to be washed-away by the sum over the infinite array.}.

Another puzzle is that in the D-brane analysis taking a periodic array leads to an infinite tower of massive states. These have a natural interpretation in string theory as the Kaluza-Klein (KK) modes of the T-dual D-brane that is wrapped on a circle. But when taking such an array in M-theory one would not expect to find an extra tower of KK-like states of D2-branes. What happens to these modes?

Nevertheless, even with all these difficulties, since the ABJM theory  is supposed to describe an arbitrary number of M2-branes, and at least for large $k$ it is weakly coupled and perturbatively reliable, there ought to be some prescription for studying the periodic array and obtaining a suitable description of D2-branes, and more generally D$p$-branes,  from M-theory. The aim of this paper then is to do just that. We  note that there are also other papers that relate M2-branes to D$p$-branes  \cite{Ho:2009nk,Kobo:2009gz,Honma:2009bx,Nastase:2010ft}.

Another motivation for studying arrays of M2-branes is that one might expect that a cubic periodic array of M2-branes could somehow be related via an M-theory version of T-duality, to M5-branes wrapped on ${\mathbb T}^3$.

The rest of this paper is organized as follows. In section two we discuss the M2-brane array and the way in which we impose discrete translational invariance on the ABJM theory and the regularization method that we use. In section three we then evaluate the Lagrangian to obtain the Lagrangian for the periodic M2-brane array. In section four we show how this Lagrangian is related to that of three-dimensional maximally supersymmetric Yang-Mills and hence D2-branes in string theory. In section five we consider a further T-duality along a transverse torus which maps our result to a non-manifest-Lorentz invariant five-dimensional Lagrangian which is similarly related to five-dimensional maximally supersymmetric Yang-Mills. Finally in section six  we give our conclusions.


\section{Set-up}

The ABJM Lagrangian is
\be
{\cL}_{ABJM}=-\Tr(D_{\mu}Z^{A}D^{\mu}\bar{Z}_{A})-i\Tr (\bar\Psi^{{A}}\gamma^\mu D_\mu\Psi_{{A}})+\cL_{Yukawa}-V+\cL_{CS}\,,
\ee
where
\be\ba{cll}
D_{\mu}Z^{A}&=&\partial_{\mu}Z^{A}-iA^{L}_{\mu}Z^{A}+iZ^{A}A^{R}_{\mu}\,,\\[6pt]
V&=&-\frac{2}{3}\Tr\left([Z^{A},Z^{B};\bar{Z}_{C}][\bar{Z}_{A},\bar{Z}_{B};Z^{C}]-\frac{1}{2}[Z^{A},Z^{B};\bar{Z}_{A}][\bar{Z}_{C},\bar{Z}_{B};Z^{C}]\right) \, , \\[6pt]
\cL_{Yukawa} &=&  - i\Tr(\bar\Psi^A[\Psi_A,Z^B;\bar Z_B])+ 2i\Tr(\bar\Psi^A[\Psi_B,Z^B;\bar Z_A])\\&&+\frac{i}{2}\varepsilon_{ABCD}\Tr(\bar\Psi^A[ Z^B,  Z^C;\Psi^D])-\frac{i}{2}\varepsilon^{ABCD}\Tr(\bar Z_D [\Psi_A,\Psi_B;\bar Z_C]) \, , \\[6pt]
\cL_{CS}&=&\frac{k}{4\pi}\varepsilon^{\mu\nu\lambda}\left(\Tr( A^{L}_{\mu}\partial_{\nu} A^{L}_{\lambda}-\frac{2i}{3} A^{L}_{\mu} A^{L}_{\nu} A^{L}_{\lambda}) -\Tr( A^{R}_{\mu}\partial_{\nu}A^{R}_{\lambda}-\frac{2i}{3} A^{R}_{\mu} A^{R}_{\nu} A^{R}_{\lambda}) \right)\,,
\ea
\ee
and
\be
[Z^A,Z^B;\bar Z_C] = \frac{2\pi}{k}(Z^A\bar Z_CZ^B -Z^B\bar Z_CZ^A )\ .
\ee

At this point we should mention our
conventions. Firstly $A,B=1,2,3,4$, $A',B'=1,2,3$, $\mu = 0,1,2$ and $\bar\Psi_A = \Psi^T_A\gamma_0$, where $\gamma_\mu$ are a real basis of the three-dimensional Clifford algebra.  We raise/lower the $SU(4)$ $A,B$ indices when taking a hermitian conjugate. To describe an infinite array of M2-branes we need to consider infinite matrices $Z^A_{mn}$, where $m,n\in \mathbb Z$ and,  for each $m,n$, $Z^A_{mn}$ is itself an $M\times M$ matrix. We use  $\bar Z_A$ to denote the hermitian matrix conjugate of the full infinite dimensional system: in components $(\bar Z_A)_{mn}=Z^\dag_{A \, nm}$, where $\dag$ denotes the matrix hermitian conjugate of the internal $M\times M$ matrix.

The maximally supersymmetric vacua of this Lagrangian consist of commuting scalars. Hence the configuration (\ref{config}) is indeed a good vacuum and describes $M$ M2-branes located at ${{\rm Im} \, Z^4 = 2\pi i n}$ for every $n\in{\mathbb Z}$. The infinite array is invariant under the shift symmetry $Z^4\to Z^4 + 2\pi i R $:
\begin{eqnarray}
\langle Z^4_{mn}\rangle &\to& \langle Z^4_{mn}\rangle + 2\pi i R \delta_{mn} \nonumber \\
&=& 2\pi i R (n+1)\delta_{mn} \\
&=& \langle Z^4_{m+1 \, n+1}\rangle\nonumber\ .
\end{eqnarray}
Next we need to impose the above finite shift symmetry on the whole theory, including the fluctuations.
 We can think of this as an orbifold action on M2-branes where the orbifold group is $\Gamma = \mathbb Z$ acting by
 \be\label{orbifold}
  Z^A_{mn}\to  Z^A_{m+1 \, n+1}\ ,
 \ee
 and similarly for the other fields.
 We must then consider configurations of M2-branes that are invariant under the action of $\Gamma$ , with the exception of $Z^4$ which is allowed to
carry integer `winding number' along the array:
 \be\ba{l}
Z^{A}_{mn}=2\pi i R\delta^{A}_{4}\delta_{mn}+Z^{A}_{(m-1)(n-1)}\,,\\
A^{L}_{\mu\,mn}=A^{L}_{\mu\,(m-1)(n-1)}\,,~~~~~~A^{R}_{\mu\,mn}=A^{R}_{\mu\,(m-1)(n-1)}\,,\\
\Psi_{A \, mn} = \Psi_{A \, (m-1)(n-1)}\ .
\ea \ee
As mentioned above the problem with this group action is that it is not a symmetry of the Lagrangian. Imposing it leads to additional constraints. Furthermore it is not consistent with the  supersymmetry transformations. Nevertheless we simply proceed and  consider the theory in this case.

We first note that the infinite size of the array leads to divergent terms in the Lagrangian. For example consider the kinetic term for the scalars\begin{eqnarray}
\sum_{m,n} \tr (\partial_\mu Z^A_{mn} \partial^\mu \bar Z_{A \,nm} )&=&
\sum_{m,n} \tr(\partial_\mu Z^A_{m-n 0 }\partial^\mu Z^\dag_{A \, m-n 0}) \nonumber \\
&=& \sum_q \sum_p \tr(\partial_\mu Z^A_{p 0}\partial^\mu Z^\dag_{A\,p 0}) \\
&=& |\Gamma| \sum_p \tr(\partial_\mu Z^A_{p 0}\partial^\mu Z^\dag_{A\,p 0}) \ , \nonumber
\end{eqnarray}
where
\be
|\Gamma| = \sum_q 1\ .
\ee
In the D-brane case \cite{Taylor:1996ik} the effect of this divergence is harmless as each term in the Lagrangian comes with the same overall factor of $|\Gamma|$. In our case however,
the fact that the shift invariance we impose is not a symmetry of the Lagrangian, causes other divergences to appear. We therefore need a way to regulate and compare divergences.

To do this we simply consider a very large but finite array consisting of M2-branes located at $Z^4 = 2\pi in R$ with $n=-N,...,N$. We then always impose the limit $N\to \infty$ in any final expressions and therefore only consider the leading large $N$ terms. Using this regulator we see that
\begin{equation}
|\Gamma| = \sum_q 1  = 2N+1 \sim 2N\ ,
\end{equation}
where $\sim$ denotes the leading order behaviour as $N\to\infty$.
We will also be cavalier about ignoring possible boundary effects that occur when $N$ is finite.  Thus our starting point is a ${U((2N+1)M)\times U((2N+1)M)}$ ABJM model with $N>>1$. Note that in such a theory the 't Hooft coupling constant grows as $NM/k$. Furthermore, when taking the limit $N\to \infty$, we will allow for both $k$ and  $R$  to scale in appropriate ways with $N$.

With this in mind we note that we can solve the shift symmetry condition in terms of  the $M\times M$ matrix valued fields $\phi^A_{n}$, $\psi_{A\,n}$, $a^{L/R}_{\mu\, n}$:
\begin{equation}
Z^{4}_{mn}:= 2\pi i R n 1_{M\times M}\delta_{mn}+\frac{1}{\sqrt{2N}}\phi^4_{n-m} \ , \quad  Z^{A'}_{mn}:=\frac{1}{\sqrt{2N}}\phi^{A'}_{n-m} \ , \nonumber
\end{equation}
\begin{equation}
A^{L/R}_{\mu\,mn}:=a^{L/R}_{\mu\,n-m} \ , \quad\Psi_{A \, mn} := \frac{1}{\sqrt{2N}}\psi_{A \, n-m}\ .
\end{equation}
Here we have included factors of $(2N)^{-\frac{1}{2}}$ so that the fields $\phi^A_p$ and $\psi_{A \, p}$ have canonical kinetic terms.
Note that since $A^{L/R}_\mu$ are hermitian we require that  $(a^{L/R }_{\mu\, n})^\dag = a^{L/R }_{\mu \,-n}$. We have not rescaled the gauge fields by $(2N)^{-\frac{1}{2}}$ since their role in covariant derivatives and gauge field strengths does not readily allow for this.


\section{Reduced Lagrangian}

Having set up our configuration we can now construct the reduced action for the infinite array. Let us start with the kinetic terms:
\begin{align}
\nonumber
D_{\mu}Z^{4}_{mn}=&\frac{1}{\sqrt{2N}}\nabla_{\mu}\phi^{4}_{n-m} -\frac{i}{\sqrt{2N}}\sum_{p\ne 0}[\,a^{+}_{\mu\,p}\,,\phi^{4}_{n-m-p}\,]-\frac{i}{\sqrt{2N}}\sum_{p}\{a^{-}_{\mu\,p}\,,\phi^{4}_{n-m-p}\} \\
\nonumber
&+2\pi R(n-m)a^{+}_{\mu\,n-m}+2\pi R(n+m) a^{-}_{\mu\,n-m}\ ,\\[11pt]
D_{\mu}Z^{A'}_{mn}=&\frac{1}{\sqrt{2N}}\nabla_{\mu}\phi^{A'}_{n-m}-\frac{i}{\sqrt{2N}}\sum_{p\ne 0}[\,a^{+}_{\mu\,p}\,,\phi^{A'}_{n-m-p}\,]-\frac{i}{\sqrt{2N}}\sum_{p}\{a^{-}_{\mu\,p}\,,\phi^{A'}_{n-m-p}\}\ , \label{mass} \\[11pt]
\nonumber
D_{\mu}\Psi_{{A}\,mn}=&\frac{1}{\sqrt{2N}}\nabla_{\mu}\psi_{{A}\,n-m}-\frac{i}{\sqrt{2N}}\sum_{p\ne 0 }[\,a^{+}_{\mu\,p}\,,\psi_{{A}\,n-m-p}\,]-\frac{i}{\sqrt{2N}}\sum_{p}\{a^{-}_{\mu\,p}\,,\psi_{{A}\,n-m-p}\} \ ,
\end{align}
where $a^{\pm}_{\mu\,n}:=\half(a^{L}_{\mu\, n}\pm a^{R}_{\mu\,n})$ and
\be
\nabla_\mu \phi^A_p = \partial_\mu\phi^A_p - i[a^+_{\mu \,0},\phi^A_p]\ .
\ee
Note the appearance of terms involving $m + n$ on the right-hand-side of \eref{mass}. These arise because the Lagrangian is not invariant under our orbifold action (\ref{orbifold}). This leads to a divergent term of the form
\begin{eqnarray}\label{div1}
(2\pi R)^2\sum_{m,n}(m+n)^2 \tr ( a^{-}_{\mu\,n-m}a^{\mu -}_{\,m-n} ) &=& (2\pi R)^2\sum_q \sum_p (p+2q)^2\tr ( a^{-}_{\mu\,p}a^{\mu -}_{\,-p} ) \nonumber \\  &\sim &\frac{8}{3}N^3(2\pi R)^2 \sum_p  \tr ( a^{-}_{\mu\,p}a^{\mu -}_{\,-p} ) \ .
\end{eqnarray}
This diverges (unless $R$ is taken to vanish at least as fast as $N^{-\frac{3}{2}}$, which we will not consider here) and is not cancelled by anything else in the Lagrangian. We therefore conclude that, to obtain finite energy configurations we must have
\be\label{aconstraint}
0 = a^{-}_{\mu\,p} =\frac{1}{2} (a^{L}_{\mu\,p}-a^{R}_{\mu\,p}) \ .
\ee
Note that one might be tempted to simply rescale $a^{-}_{\mu \,p}$ by a factor proportional to $N^{-3/2}$ so as to render (\ref{div1}) finite. However one would then simply find that, in the limit $N\to\infty$, $a^-_{\mu \,p}$ drops out from   the covariant derivative and hence the  Lagrangian.
Thus (\ref{aconstraint}) should be viewed as a constraint on the system that breaks the gauge group to $U(M)$.

In particular the gauge group associated to the zero mode is just $U(M)$. This should be viewed as a constraint on the system.

Next we look at the quadratic terms that come from expanding the potential:
\begin{eqnarray}
V_{\phi^2} &=&  \frac{1}{2N}\left(\frac{2\pi}{k}\right)^2(2\pi R)^4\sum_{p,q}p^{2}(p+2q)^{2} \tr(\phi^{A'}_{p}{\phi}^\dag_{A' \,p}) \nonumber \\
&\sim &  M_b^2\sum_{p}p^{2} \tr(\phi^{A'}_{p}{\phi}^\dag_{A' \,p}) \ ,
\end{eqnarray}
where
\be
M_b^2 = \frac{1}{2N}\left(\frac{2\pi}{k}\right)^2(2\pi R)^4 \sum_q (p+2q)^2\sim \frac{4}{3}(2\pi)^6\frac{N^2R^4}{k^2}\ .
\ee
Although there could be cases where it is finite if $R\to 0$ sufficiently quickly, we will consider the case that $M_b\to \infty$ as $N\to \infty$.
Here we see that the mysterious KK-like tower is lifted to infinite mass, resolving one of the puzzles raised in the introduction. In particular
we must impose the constraint:
\be
\phi^{A'}_p =0\ , \qquad p\ne 0\ .
\ee
Note that the masslessness of $\phi^4_p$ does  not seem related to the problem mentioned in the introduction, where spurious massless states arise when pairs of M2-branes are at equal distance from the origin, since that degeneracy applies to all four scalars in the same way.

Let us next examine the quadratic fermion term:
\begin{align}\label{fmass}
{\cal L}_{\psi^2}=& \ \frac{i}{2N}\left(\frac{2\pi}{k}\right)(2\pi R)^2\sum_{p,q} (p^2+2pq) \tr ( \bar\psi^{A'}_p   \psi_{A'\,p} ) -\frac{i}{2N} \left(\frac{2\pi}{k}\right)(2\pi R)^2\sum_{p,q} (p^2+2pq) \tr (\bar\psi^{4}_p   \psi_{4\,p} ) \ \nonumber\\
\sim & \ iM_f\sum_p p \tr ( \bar\psi^{A'}_p \psi_{A'\,p} ) -iM_f\sum_p p \tr ( \bar\psi^{4}_p   \psi_{4\,p} ) \ ,
\end{align}
where
\be
M_f = \frac{1}{2N}\left(\frac{2\pi}{k}\right)(2\pi R)^2\sum_q (p+2q)  \sim \Omega(2\pi)^3\frac{R^2N}{k}\ .
\ee
Here we have used the regularization
\be
\sum_q q \sim \Omega N^2\ ,
\ee
where $\Omega$ is an undetermined constant of order 1. In particular we note that this  sum is ill-defined. To determine how to treat it we will  use supersymmetry.\footnote{It is conceivable that this ambiguity can be avoided by performing our calculations with a superspace formalism.} This suggests that the fermion masses should be the same as the bosons, {\it i.e.} $M_f=M_b$, and hence gives
\be
\Omega = {\frac{2}{\sqrt3}}\ ,
\ee
however we will keep $\Omega$ general in our calculations in this section.
Assuming $\Omega\ne 0$ we conclude that there is also a fermionic constraint
\be
\psi^A_{p}=0\ , \qquad p\ne 0.
\ee

Thus  we see that in order to avoid divergences in the Lagrangian (and correspondingly Hamiltonian) we must impose the constraints
\be
\phi^{A'}_p=0\ , \qquad \psi_{A\,p}=0\ ,\qquad a^{-}_{\mu \,0 }=0\ ,\qquad a^{-}_{\mu \,p }=0\ ,\qquad p\ne 0\ ,
\ee
where $A'=1,2,3$. This leaves us with the zero-modes
\be\label{zeromodes}
\phi^A_0\ ,\qquad \psi_{A\,0}\ ,\qquad a^+_{\mu\, 0}\ ,
\ee
as well as three infinite towers of fields:
\be
a^+_{\mu \,p}\ ,\qquad \chi_p = \frac{1}{2} \phi^4_p + \frac{1}{2} \phi^\dag_{4\,-p}\ ,\qquad \omega_p = -\frac{i}{2}\phi^4_p +  \frac{i}{2}\phi^\dag_{4\,-p} \ ,\qquad p\ne 0 \ ,
\ee
which satisfy  $\omega^\dag_p = \omega_{-p}$, $\chi^\dag_p = \chi_{-p}$ and $(a^+_{\mu \,p})^\dag = a^+_{\mu \,-p}$.

Once we have set these infinitely massive fields to zero we must also ensure that there are no source terms for them in the action. Classically this is a clear requirement to solve the equations of motion. Quantum mechanically it follows from the general procedure for quantization with a constraint. In particular if we have a Hamiltonian $H(\vec q,\vec p)$ on some phase space with coordinates $(\vec q,\vec p)$ and impose a constraint $C(\vec q,\vec p)=0$ then we require that $\{H,C\}=0$ so that the constraint is consistent with time evolution. In general this leads to a new set of constraints $\{H,C\} = C_1$, $\{H,C_1\}=C_2$ {\it etc.} In our case the original constraints simply set $\phi^{A'}_p=a^-_{\mu \,p}=\psi_{A\,p}=0$. One then finds that the resulting additional constraints are simply the vanishing of the sources for $\phi^{A'}_p, a^-_{\mu\, p}$ and $\psi_{A\,p}$, $p\ne 0$. We take the view here that, in order to ensure a smooth large $N$ limit,  such sources must be set to zero even for finite, but large $N$. In addition this means that sources that scale differently with $N$ must be made to vanish separately. However we don't expect that our results depend significantly on this.

Let us examine such sources. First  we look at the kinetic terms. Here we see that there will be a source for $\phi^{A'}_p$, $p\ne 0$ arising from $a^+_{\mu\, p}$:
\be
D^2 Z^{A'}_{mn} = \frac{1}{\sqrt{2N}}\partial^2 \phi^{A'}_0\delta_{mn}- \frac{i}{\sqrt{2N}}[\partial_\mu a^{\mu +}_{n-m},\phi^{A'}_{0}]- \frac{2i}{\sqrt{2N}}[ a^{\mu +}_{n-m},\partial_\mu \phi^{A'}_{0}] - \frac{1}{\sqrt{2N}}\sum_{p}[a^+_{\mu\, p-m},[a^{\mu +}_{n-p},\phi^{A'}_{0}]] .
\ee
Thus we also require that $a^+_{\mu\, p}$ is proportional to the $M\times M$ identity matrix if $p\ne 0$. This means that $a^+_{\mu\, p}$, $p\ne 0$, does not appear in $D_\mu Z^{A'}_{mn} = \frac{1}{\sqrt{2N}}\nabla_\mu \phi^{A'}_0\delta_{mn}$.

We can also expand the potential to cubic order in $\phi^A_p$. Although this vanishes if $\phi^{A'}_p=0$, $p\ne 0$, one does find a source term for $\phi^{A'}_p$, $p\ne 0$:
\begin{eqnarray}
V_{\phi^3} &=&i\frac{(2 \pi R)^{3}}{(2N)^{3/2}}\left(\frac{2\pi}{k}\right)^2 \sum_{p,q}\tr\left[\left(p^2(p+2q)\{\chi_{p},\phi^{A'}_{0}\}+ip(p+2q)^2[\omega_{p},\phi^{A'}_{0}]\right)\phi_{A'\, p}\right]+h.c. +\ldots\nonumber\\
&\sim & i\frac{(2 \pi )^{5}}{\sqrt{2}}\sum_{p}\tr\left[\left(\Omega\frac{N^{1/2}R^3}{k^2}p^2\{\chi_{p},\phi^{A'}_{0}\}+i\frac{4N^{3/2}R^3}{3k^2}p[\omega_{p},\phi^{A'}_{0}]\right)\phi_{A'\, p}\right]+h.c.+\ldots\ ,
\end{eqnarray}
where the ellipsis denote further cubic terms that  are not linear in $\phi^{A'}_p$ and hence not sources. Requiring that this vanishes tells us that
\be
\chi_p=0\ ,\qquad \omega_p \propto 1_{M\times M} \ , \qquad  p\ne 0\ .
\ee

There is also a source for $a^-_{\mu\, p}$. Setting this to zero leads to the constraint\footnote{{We denote $[\bar\Psi^{A},\gamma_\mu\Psi_{A}]:=\bar\Psi^{Aa}\gamma_\mu\Psi_{Ab}[T^\dagger_a,T^b]$.}}
\be
[Z^A,D_\mu  \bar Z_A] +  [\bar Z_A,D_\mu  Z^A] - i [\bar\Psi^A,\gamma_\mu\Psi_A]=0\ . 
\ee
The non-zero mode part of this constraint leads to
\be
2\pi i R p a^{+}_{\mu\, p} =\frac{1}{\sqrt{2N}}\nabla_{\mu}\omega_{p} \ . \label{efg}
\ee
The zero-mode part of the constraint is the $a^+_{\mu\, 0}$ equation of motion and will be dealt with later. One then finds that, substituting back into the Lagrangian,  (\ref{efg}) simply removes both $\omega_p$ and $a^+_{\mu\, p}$  all together.

Thus, once all the constraints are fully considered we are effectively left with just the zero-modes (\ref{zeromodes}).
We can now evaluate the Lagrangian. In the limit that $N\to\infty$ the sixth order terms in the potential and fourth order terms in the Yukawa interaction vanish. The final result for the Lagrangian evaluated on an infinite M2-brane array is
\begin{align}\label{M2act}
{\cal L}_{array} =& -\tr(\nabla_\mu\phi^{A'}_0\nabla^\mu\phi^\dag_{A'\, 0}) -\tr(\nabla_\mu{\rm Re} \, \phi^4_0\nabla^\mu{\rm Re} \, \phi^4_0) -\tr(\nabla_\mu{\rm Im} \, \phi^4_0\nabla^\mu{\rm Im} \, \phi^4_0)- i\tr(\bar\psi^A_0\gamma^\mu\nabla_\mu \psi_{A\,0})\nonumber\\ &+{\cal L}_{Yukawa} -V\ ,
\end{align}
where ${\rm Re} \, \phi^4_0 = \frac{1}{2}(\phi^4_0+\phi^\dag_{4\,0})$, ${\rm Im} \, \phi^4_0 = -\frac{i}{2}(\phi^4_0-\phi^\dag_{4\,0})$. The potential and Yukawa terms are
\begin{align}
V =& - \frac{g^2_{YM}}{2} \tr\left( [\phi^{A'}_0,\phi^\dag_{B'\,0}][\phi^\dag_{A'\,0},\phi^{B'}_{0}]+[\phi^{A'}_0,\phi^ {B'}_{0}][\phi^\dag_{A'\,0},\phi^\dag_{B'\,0}]+4[\phi^{A'}_0,{\rm Im} \, \phi^4_0][\phi^\dag_{A'\,0},{\rm Im} \, \phi^4_0]\right) \ , \nonumber\\[6pt]
\nonumber
{\cal L}_{Yukawa} =& \phantom{-} g'_{YM} \tr \left( 2i\bar\psi^{A'}_0[{\rm Im} \, \phi^4_0,\psi_{A'\,0}]-2i\bar\psi^{4}_0[{\rm Im} \, \phi^4_0,\psi_{4\,0}]+2\bar\psi^{A'}_0[\phi^\dag_{A'\,0},\psi_{4\,0}]+2\bar\psi_{A'\,0}[\phi^{A'}_0,\psi^{4}_0]\right. \\
&\left. \ \qquad \qquad+\varepsilon_{A'B'C'}\bar{\psi}^{A'}_0[ \phi^{B'}_0,\psi^{C'}_0]+\varepsilon^{A'B'C'}\bar{\psi}_{A'\,0}[ \phi_{B'\,0},\psi_{C'\,0}]\right)\ ,
\end{align}
with
\begin{eqnarray}
g^2_{YM}&=& \frac{(2\pi R)^2}{2N^2} \left(\frac{2\pi}{k}\right)^2 \sum_q q^2  \sim   \frac{1}{3}(2\pi)^4 \frac{R^2N}{k^2}\ , \nonumber\\
g'_{YM} &=& \frac{ 2\pi R }{(2 N)^{3/2}}\left(\frac{2\pi}{k}\right) \sum_{q}q \sim \frac{\Omega}{2\sqrt{2}} (2\pi)^2 \frac{R \sqrt{N}}{k} =\sqrt{\frac{3}{8}}\Omega g_{YM}\ .
\end{eqnarray}
Thus to obtain an interesting theory, with Lagrangian (\ref{M2act}), we require $M_b\to\infty$ with $g_{YM}$ finite in the limit $N\to \infty$.


\section{Comparing to Three-Dimensional Maximally Supersymmetric Yang-Mills}

The theory we have obtained looks rather strange as there is no kinetic term for the gauge field. Therefore we should consider comparing our result to three-dimensional maximally supersymmetric Yang-Mills that is obtained from the open string description of D2-branes:
\begin{eqnarray}
{\cal L}_{3DMSYM} &=& -\frac{1}{4 g^2_{YM}} \tr ( F_{\mu\nu}F^{\mu\nu} ) -  \frac{1}{2}\tr ( \nabla_\mu X^I\nabla^\mu X^I ) - \frac{i}{2}\tr ( \bar\Lambda\Gamma^\mu \nabla_\mu \Lambda ) \nonumber \\
&& +\frac{g_{YM}}{2}\tr ( \bar\Lambda\Gamma^{11}\Gamma^I[X^I,\Lambda] ) +\frac{g^2_{YM}}{4}\tr \sum_{I,J}([X^I,X^J])^2\ ,
\end{eqnarray}
where $\bar \Lambda = \Lambda^T\Gamma_0$ and $\nabla_\mu X^I =\partial_\mu X^I - i[A_\mu, X^I]$.
Here there are seven scalars $X^I$, $I=3,4,..,9$, a gauge field $A_\mu$ and fermions $\Lambda$ which, as written, form a real 32-component $SO(1,9)$ spinor that satisfies $\Gamma_{012}\Lambda = -\Lambda$. Furthermore  $\Gamma_\mu$, $\Gamma_I$ are real $32\times 32$ $\gamma$-matrices and $\Gamma^{11} = \Gamma_0...\Gamma_9$.

To compare with our results we need to break the manifest $SO(7)$ symmetry to $SU(3)$. To this end we let (we will consider the fermions shortly)
\begin{equation}
X^{A'+2} = \frac{1}{\sqrt{2}}(\phi^{A'}_0+\phi^\dag_{A' \,0}) \ , \qquad X^{A'+5} = \frac{1}{\sqrt{2i}}(\phi^{A'}_0-\phi^\dag_{A' \,0}) \ , \qquad X^9 = \sqrt{2}{\rm Im} \, \phi^4_0\ ,
\end{equation}
where
\begin{equation}
{\rm Re} \, \phi^A_0 = \frac{1}{2}(\phi^A_0+\phi^\dag_{A\,0}) \ , \qquad {\rm Im} \, \phi^A_0 = - \frac{i}{2}(\phi^A_0-\phi^\dag_{A\,0})\ .
\end{equation}
The bosonic part of ${\cal L}_{3DSYM}$ can now be written as
\begin{align}
\nonumber
{\cal L}^{(b)}_{3DMSYM} =& -\frac{1}{4g^2_{YM}} \tr ( F_{\mu\nu}F^{\mu\nu} ) -  \tr ( \nabla_\mu \phi^{A'}_0\nabla^\mu \phi^\dag_{A'\,0} ) -  \tr ( \nabla_\mu {\rm Im} \, \phi^{4}_0\nabla^\mu {\rm Im} \, \phi^4_{0} ) \\
& +\frac{g^2_{YM}}{2}\tr \left([\phi^{A'}_0,\phi^{B'}_0][\phi^\dag_{A'\,0},\phi^\dag_{B'\,0}]+[\phi^{A'}_0,\phi^\dag_{B\,0}][\phi^\dag_{A\,0},\phi^{B'}_0]+4[{\rm Im} \, \phi^4_0,\phi^{A'}_0][{\rm Im} \, \phi^4_0,\phi^\dag_{A'\,0}]\right)\ .
\end{align}

Next we consider the fermions. Here we need to reduce $\Lambda$ to four, complex, two-component spinors $\psi^A$. To do this we note that the Clifford algebra can be reduced as
\be
\Gamma_\mu = \gamma_\mu\otimes \rho_\star\qquad \Gamma^{I} = 1\otimes \rho^{I-2}\ ,
\ee
where $\rho^1,...,\rho^8$ are a real, $16\times 16$-matrix representation of the Euclidean eight-dimensional Clifford algebra and $\rho_\star = \rho^1\rho^2...\rho^8$. In this formulation $\Gamma^{11} = 1\otimes\rho^8$ and $\Gamma_{012} = 1\otimes\rho_\star$. We can therefore decompose
\be
\Lambda = \lambda_\Sigma\otimes \eta^\Sigma = \frac{1}{\sqrt{2}}\left(\lambda_A\otimes \eta^A+\lambda^A\otimes \eta_A\right)\ ,
\ee
where $\Sigma=1,2,3,...,16$ and $\rho_\star\eta^\Sigma=-\eta^\Sigma$ (so that there are just eight independent $\eta^\Sigma$) which we take to be normalized such that $(\eta^{\Sigma})^T \eta^\Pi=\delta^{\Sigma\Pi}$. We have also introduced a complex basis of spinors $\eta^A$, along with a suitable complex basis of $\rho^A$-matrices,  that will be useful later.
The fermion terms are now
\begin{align}
{\cal L}^{(f)}_{3DMSYM} =& -\frac{i}{2} \tr ( \bar\lambda_\Sigma \gamma^\mu \nabla_\mu \lambda_\Pi ) ( \delta^{\Sigma\Pi} ) \nonumber \\
& -\frac{g_{YM}}{2\sqrt{2}}\tr(\bar \lambda_\Sigma [\phi^{A'}_0,\lambda_\Pi])\left((\eta^{\Sigma})^T\rho_8\rho_{A'}\eta^\Pi\right)-\frac{g_{YM}}{2\sqrt{2}}\tr(\bar \lambda_\Sigma [\phi_{A'\,0},\lambda_\Pi])((\eta^{\Sigma})^T\rho_8\rho^{A'}\eta^\Pi)\nonumber\\
& + \frac{ig_{YM}}{2\sqrt{2}}\tr(\bar \lambda_\Sigma [\phi^4_0,\lambda_\Pi])\left((\eta^{\Sigma})^Ti \rho_8\rho_{7}\eta^\Pi\right) + \frac{ig_{YM}}{2\sqrt{2}}\tr(\bar \lambda_\Sigma [\phi^\dag_{4\,0},\lambda_\Pi])\left((\eta^{\Sigma})^T i\rho_8\rho_{7}\eta^\Pi\right) \, ,
\end{align}
where $\bar\lambda_\Sigma  = \lambda^T_\Sigma\gamma_0$.

Let us first consider the last line. If we consider complex fermions then we can diagonalize $i\rho_8\rho_7$ with
\be
i\rho_8\rho_7\eta^{A'} = \eta^{A'} \, , \qquad i\rho_8\rho_7\eta_4 = \eta_4\ .
\ee
It then follows that the complex conjugates satisfy
\be
i\rho_8\rho_7\eta_{A'} = -\eta_{A'} \, , \qquad i\rho_8\rho_7\eta^4 = - \eta^4\ .
\ee
We can choose to normalize this basis such that
\be
(\eta^A)^T\eta_B = 2\delta^A_B\ ,\qquad (\eta_A)^T\eta_B = 0\ .
\ee
Next we need to deduce the action of $\rho_8\rho_{A'}$, $\rho_8\rho^{A'}$ on $\eta^{B'}$,  $\eta_4$ and their complex conjugates. We note that the Clifford algebra is equivalent to
\be\label{blah}
\{\rho_8\rho_{A'},\rho_8\rho_{B'}\} =0\ ,\qquad \{\rho_8\rho_{A'},\rho_8\rho^{B'}\} =-4\delta_{A'}^{B'}\ .
\ee
Since we haven't been very precise about the exact definition of $\eta^{A'}$ and $\eta_4$ it is enough to observe that the choice
\begin{eqnarray}
\rho_8\rho_{A'}\eta^{B'} &=& 2\delta_{A'}^{B'}\eta^4\ ,\qquad\rho_8\rho_{A'}\eta_{B'} = -2\varepsilon_{A'B'C'}\eta^{C'} \ , \nonumber\\
 \rho_8\rho_{A'}\eta_{4} &=& -2\eta_{A'}\ ,\qquad\rho_8\rho_{A'}\eta^{4} = 0\ ,
\end{eqnarray}
and similarly for the complex conjugates,
satisfies the algebra (\ref{blah}). In this basis the fermion terms become
\begin{eqnarray}
{\cal L}^{(f)}_{3DMSYM} &=& -i \tr \left( \bar\lambda^{A} \gamma^\mu \nabla_\mu \lambda_{A} \right)   \nonumber \\
&& +\frac{g_{YM}}{\sqrt{2}}\tr\left(2i\bar \lambda^{A'} [{\rm Im} \, \phi^4_{0},\lambda_{A'}]-2i\bar \lambda^{4} [{\rm Im} \, \phi^4_{0},\lambda_{4}] + 2\bar \lambda^{A'}[\phi^\dag_{A'\,0},\lambda_{4}] - 2\bar \lambda^4 [\phi^{A'}_{0},\lambda_{A'}]\right.\nonumber\\
&&\left.  +  \varepsilon_{A'B'C'}\bar \lambda^{A'} [\phi^{B'}_0,\lambda^{C'}] + \varepsilon^{A'B'C'} \bar \lambda_{A'} [\phi_{B'\,0},\lambda_{C'}]\right)\ .
\end{eqnarray}
 In particular we see that, if we identify $A_\mu = a^+_{\mu \,0}$, $\lambda^A=\psi^A_{0}$ and take $g'_{YM}=g_{YM}/\sqrt{2}$, corresponding to $\Omega = 2/\sqrt{3}$ as before, then (\ref{M2act}) can be  written as
\begin{eqnarray}
{\cal L}_{array}&=&
-\frac{1}{2} \tr \left( \nabla_\mu Y \nabla^\mu Y \right) -  \frac{1}{2}\tr \left( \nabla_\mu X^I\nabla^\mu X^I \right) - \frac{i}{2}\tr \left( \bar\Lambda\Gamma^\mu \nabla_\mu \Lambda \right) \nonumber \\
&&  +\frac{g_{YM}}{2}\tr \left( \bar\Lambda \Gamma^{11}\Gamma^I[X^I,\Lambda] \right) +\frac{g^2_{YM}}{4} \sum_{I,J} \tr ([X^I,X^J])^2\ ,
\end{eqnarray}
where
\be
Y = {\rm Re} \, \phi^4_0\ .
\ee
Thus we find that the three-dimensional maximally supersymmetric Yang-Mills Lagrangian is in agreement with the M2-brane Lagrangian, with the exception of the kinetic term of the gauge field, which is absent, along with an additional scalar field $Y$ which does not enter in the potential. In particular we see that the M2-brane Lagrangian from the infinite array has an $SO(7)$ symmetry, which is enhanced from the manifest $SU(3)$ symmetry that we started with.

Since our action has no kinetic term for the gauge field, its equation of motion imposes a constraint:
\be\label{noJ}
i[Y,\nabla_\mu  Y] +   i[X^I,\nabla_\mu  X^I]+\frac{1}{2}[\bar\Lambda , \Gamma_\mu\Lambda]=0\ .
\ee
Furthermore the scalar $Y$ couples to the gauge field but does not enter into the potential. Its equation of motion is
\be
\nabla^2Y=0\ .
\ee
A solution to this equation is
\be\label{dualization}
\nabla_\mu Y = -\frac{1}{2 g_{YM}}\varepsilon_{\mu\nu\lambda}F^{\nu\lambda}\ .
\ee
From this we deduce that
\begin{eqnarray}
[Y,\nabla_\mu  Y] &=&- \frac{1}{2 g_{YM}}\varepsilon_{\mu\nu\lambda}[Y,F^{\nu\lambda}]\nonumber\\
&=& \frac{i}{ g_{YM}}\varepsilon_{\mu\nu\lambda}\nabla^\nu\nabla^\lambda Y\nonumber\\
&=&\frac{i}{ g^2_{YM}}\nabla^\nu F_{\mu\nu}\ .
\end{eqnarray}
In which case we find that the  constraint (\ref{noJ}) can be written as
\be
\frac{ 1}{ g^2_{YM}}\nabla^\nu F_{\mu\nu} = i[X^I,\nabla_\mu  X^I]+\frac{1}{2}[\bar\Lambda, \Gamma_\mu\Lambda]\ ,
\ee
which is precisely the equation for three-dimensional maximally supersymmetric Yang-Mills. In particular we have recovered all 16 supersymmetries in addition to the $SO(7)$ R-symmetry. Note that this is an on-shell dualization of the scalar field into a gauge field. Without this dualization ${\cal L}_{array}$ is not supersymmetric, however one may conjecture that it secretly enjoys a hidden quantum supersymmetry, much like the case of the enhanced maximal supersymmetry in the ABJM models at $k=1,2$ (although here it would seem to appear even at weak coupling).

Thus the system we obtained is related classically to 3D-SYM. In particular, to be more precise, every solution to 3D-SYM solves our system. However our system is slightly more general. In particular consider a pure-gauge configuration
\be
A_{\mu  } = i g\partial_\mu g^{-1}\ ,
\ee
where $g\in U(M)$.
The solution to (\ref{dualization}) is
\be
Y = gY_0 g^{-1}\ ,
\ee
where $Y_0$ is any constant hermitian matrix. Thus there is an additional, non-dynamical, `modulus' that appears in the M2-brane description of D2-branes. This is not surprising and should be thought of as the positions of the D2-branes in the eleventh dimension. In particular it is possible to break the gauge group while keeping all the D2-branes at the origin of the string theory Coulomb branch $X^I=0$. Note that the classical vacuum moduli space condition does not require that the vevs of $Y$ and $X^I$ commute.

Let us now discuss some curiosities of our results. We see that $X^9=\sqrt{2}{\rm Im} \, \phi^4_0 $ appears in the potential whereas in the original array ${\rm Im} \, Z^4$ represents the direction along the array, {\it i.e.}\ the M-theory direction, and is subject to a discrete shift symmetry: ${\rm Im} \, Z^4 \to {\rm Im} \, Z^4 +2\pi  R $. This symmetry is still present in our system at finite $N$,  although due to the field normalization it is rescaled to ${\rm Im} \, \phi^4_0 \to {\rm Im} \, \phi^4_0 +2\pi \sqrt{2N} R$ which diverges when we take $N\to\infty$. However in the D-brane interpretation $X^9$ is not the M-theory direction.

As mentioned above we need to take a limit $N\to\infty$ such that $M_b\to\infty$ and $g_{YM}$ finite. This can be done in a variety of ways. In particular since $M_b\propto g^2_{YM} k$ all  we require is that $N,k\to \infty$ with $R\propto k/\sqrt{N}$. We could achieve this by keeping $R$ fixed and $N\propto k^2$. Since the scalar fields have canonical dimensions of $(mass)^{1/2}$ the physical radius of the array is
\be
R_{11} = RT^{-1/2}_{M2}\ .
\ee
Since $T_{M2} = (2\pi)^{-2} l_p^{-3}$, $l_s=g_s^{-1/3}l_p$ and $R_{11}=g_sl_s$ we see that
\be
g^2_{YM} = \frac{(2\pi)^2}{3} \frac{N}{ k^2}\frac{g_s}{l_s}\ .
\ee
whereas the precise relationship for a D2-brane is $g^2_{YM} = g_s/l_s$. We can arrange for this by taking $N=ab$, $k=b$ where $a/b$ is a rational approximation to $3/(2\pi)^{2}$. However this seems very ad hoc.

One way to avoid any conflicts with these issues is to consider a scaling limit where $R\propto \sqrt{N}$, $k\propto N$. In this way we can remain at weak 't Hooft coupling $NM/k$ throughout. The distance between the M2-branes then diverges so that the fluctuations of ${\cal L}_{array}$ really just describe an isolated block of $M$ M2-branes, inside the array. (In fact this is true more generally as the normalization of $\phi^A$ ensures that  fluctuations of $\phi^A_0$ do not correspond to finite fluctuations of $Z^A$.) In this case the reduction to type IIA string theory arises because of $k\to\infty$, and the associated spacetime ${\mathbb C}^4/{\mathbb Z}_k$ orbifold, in addition to the periodicity imposed by the array.

Finally we note that it is not clear how to quantize the action we have found. Although we have derived it from the ABJM model which does have a well-defined quantization. One way is to map it to an equivalent classical Lagrangian which is more suitable to quantization, {\it i.e.}\ one which admits a simple Hamiltonian without constraints, or with constraints that can be readily solved. Ignoring the subtlety that we have mentioned above this would lead to 3D maximally supersymmetric Yang-Mills and its familiar quantization. Another approach would be to use Dirac quantization applied to the constraint induced from the $A_{\mu }$ equation of motion. We will not address this problem in this paper. Assuming that there is a suitable quantum theory involving $Y$ we can consider operators such as
\be
{\cal M} = e^{iY }\ ,
\ee
which correspond to monopole (or 't Hooft) operators. Thus we have arrived at a more refined version of three-dimensional maximally supersymmetric Yang-Mills as the description of D2-branes in type IIA string theory.

\section{Further Compactification on ${\mathbb T}^2$ and M5-branes}

Let us now consider a doubly periodic array in the $X^3$ and $X^4$ directions. For simplicity we will only consider the bosonic part of the action in this section. The extension to include the fermions is straightforward. Firstly let us rescale the scalars by a factor of $g^{-1}_{YM}$ to cast the (bosonic) action as
\be\label{bM2act}
S^{(b)}_{array}=-\frac{1}{g^2_{YM}}\tr\, \int d^3 x\
 \frac{1}{2}  \nabla_\mu Y \nabla^\mu Y  +  \frac{1}{2}  \nabla_\mu X^I\nabla^\mu X^I -\frac{1}{4} \sum_{I,J}([X^I,X^J])^2\ .
\ee
If we impose the on-shell dualization discussed above then we arrive at the familiar three-dimensional maximally supersymmetric Yang-Mills theory of D2-branes. In this case imposing a further periodic array along $X^3$, $X^4$ was studied some time ago in \cite{Taylor:1996ik} and leads to the same action as 5D maximally supersymmetric Yang-Mills compactified on ${\mathbb T}^2$. We follow the same steps here but without the dualization.

We first consider an infinite parallel array along the $X^3$ direction by imposing the shift symmetry:
\begin{eqnarray}
X^I_{mn} &=& 2\pi R' n\delta_{m,n}\delta^I_3 + X^I_{n-m} \nonumber\\
Y_{mn} &=&  Y_{n-m} \nonumber\\
A_{\mu\, m n} &=&  A_{\mu\, n-m}
\end{eqnarray}
here $m,n\in \mathbb Z$ and as before each field is an $M\times M$ hermitian matrix. Note that $R'$ has dimensions of mass. (In the interests of not introducing more symbols we are being rather brief in our notation.)  We can then repackage these fields in terms of a higher dimensional gauge theory on ${\mathbb R}^{3}\times S^1$:
\begin{eqnarray}
X^I &=&\sum_n e^{i n x^3R'}X^I_{n}\ , \quad I = 4,5,...,9 \nonumber\\
Y &=&  \sum_n e^{i n x^3R'}Y_{n}\nonumber\\
A_{\mu} &=& \sum_n e^{i n x^3R'}A_{\mu\, n}\nonumber\\
A_{3}&=&\sum_n e^{i n x^3R'}X^3_{n}\ ,
\end{eqnarray}
where $x^3$ is periodic with period   $2\pi/R'$.
Next we repeat this procedure for an array along $X^4$, with the same periodicity. In this way we construct five-dimensional hermitian matrix valued fields $Y,A_{\mu'}, X^{I'}$ where $\mu'=0,1,...,4$ and $I' =5,6,...,9$.

Following the analysis of \cite{Taylor:1996ik} leads to the action\footnote{We could also use our regularization technique of introducing a large but finite array of size $N>>1$. This would simply result in an additional factor of $4N^2$ in front of the action which we would then remove by an appropriate rescaling of $g_{YM}$ and $R'$.}
\begin{align}\label{M5act}
\nonumber
S^{(b)}_{cubic\ array}=-\frac{R'^2}{(2\pi)^2 g^2_{YM}}\tr\, \int d^5  x  \ \frac{1}{2}&  \nabla_{\mu'} X^{I'}\nabla^{\mu'} X^{I'} -\frac{1}{4} \sum_{I',J'}([X^{I'},X^{J'}])^2
 \\ 
+\frac{1}{2}& \nabla_{\mu} Y \nabla^{\mu} Y  + \frac{1}{2}F_{\mu \alpha}F^{\mu\alpha }+ \frac{1}{4}F_{\alpha\beta}F^{\alpha\beta}\ .
\end{align}
where $\alpha,\beta=3,4$ and $\mu,\nu = 0,1,2$.
This is not five-dimensional Lorentz invariant. In particular there is no kinetic term for $Y$ along the torus directions and no $F_{\mu\nu}$ terms. The first issue  arises because there is no $[X^I,Y]$ term in (\ref{bM2act}) whereas  the second  arises because there was no $F_
{\mu\nu}$ term to start with.
Nevertheless we note that the $A_\alpha$ equation of motion is that of five-dimensional maximally supersymmetric Yang-Mills:
\be
\nabla^{\mu'} F_{\alpha\mu'}  = i[X^I,\nabla_\alpha X^I]\ .
\ee
One the other hand the $A_\mu$ equation of motion is similar to before but with an extra term:
\be
\nabla^\alpha F_{\nu\alpha} =  i[Y,\nabla_\nu Y] + i[X^I,\nabla_\nu X^I]\ .
\ee
Once again we can consider the on-shell dualization and choose:
\be
\nabla_\mu Y =  -\frac{1}{2}\varepsilon_{\mu\nu\lambda}F^{\nu\lambda}\ .
\ee
which is still consistent with the $Y$ equation of motion $\nabla_\mu \nabla^\mu Y =0$. In this way we obtain
\be
\nabla^{\mu'} F_{\nu\mu'} =   i[X^I,\nabla_\nu X^I]\ ,
\ee
so that our equations are those of (the bosonic part of) five-dimensional maximally supersymmetric Yang-Mills, which restores five-dimensional Lorentz symmetry and 16 supersymmetries.

We can also consider another on-shell dualization, which is more naturally associated with the broken Lorentz symmetry, and take
\be\label{Yagain}
\nabla_\mu Y= \frac{1}{2}\varepsilon^{\alpha\beta}H_{\mu \alpha\beta}\ .
\ee
We can then also write $F_{\mu \alpha}=H_{\mu\alpha 5}$ and $F_{\alpha\beta}=H_{\alpha\beta 5}$ which is sufficient to determine all the components of a self-dual six-dimensional 3-form $H$. In this language the `moduli' $Y_0$ associated to solving (\ref{Yagain}) can be thought of as the  period of a two-form potential $B_{34}$ integrated over the two-cycle of the torus.


\section{Conclusions}

In this paper we have investigated periodic arrays of M2-branes using the ABJM model. By introducing a regularization method, where we consider a large but finite array with $2N+1$ sites, imposing a discrete translational symmetry, computing the action, and then letting $N\to\infty$. The Chern-Simons level $k$ and array radius $R$ were also allowed to scale with $N$ and to obtain a suitable theory of D2-branes we required that  $k\to\infty$ with $R\propto k/ \sqrt{N}$. Our result is a curious variation of three-dimensional maximally supersymmetric Yang-Mills where the gluon kinetic term is replaced by that of a dual scalar field. All solutions of three-dimensional maximally supersymmetric Yang-Mills can be made solutions of our Lagrangian but in addition we find  non-dynamical moduli. We also considered further doubly-periodic arrays that map the D2-branes to D4-branes. This led to a non-Lorentz invariant version of five-dimensional maximally supersymmetric Yang-Mills. This in turn can be viewed as the M-theory description of a cubic array of M2-branes, which should therefore also describe M5-branes wrapped on ${\mathbb T}^3$. 

We should mention the relation of this work to the proposal that the $(2,0)$ theory of M5-branes on $S^1$ is exactly five-dimensional maximally supersymmetric Yang-Mills\cite{Douglas:2010iu,Lambert:2010iw}. In this paper we have shown that the ABJM model  can be used to describe a cubic periodic array of M2-branes and yields an action which is essentially the same as that of five-dimensional maximally supersymmetric Yang-Mills but with additional non-dynamical `moduli'. Although we were required to rescale $k\to\infty$ to obtain this action we did keep all eleven-dimensional momentum modes. In particular the on-shell dualization of $Y$ implies that magnetic flux $F_{12}$ plays the role of the `missing' eleven-dimensional momentum. Furthermore the resulting theory has a coupling constant that we could tune to be small but which we could also take to be large. Therefore our results appear to be in broad agreement with the proposal of \cite{Douglas:2010iu,Lambert:2010iw}.

However it is not entirely clear how much our results should be trusted at strong coupling. In particular the M2-brane physics relies crucially on 't Hooft (monopole) operators which we have not addressed. Although these are not expected to be   important at large $k$ our analysis is largely justified by the fact that our results can be mapped to the known open string description obtained at weak $g_{YM}$. Therefore it remains  possible that 't Hooft operators  are important here at large $g_{YM}$.

We could also try to consider other cases such as those with both $M_b$ and $g_{YM}$ finite, which we would find if $k$ is held fixed and $R\propto N^{-1/2}$ in the limit $N\to\infty$. In this case we there is a Kaluza-Klein-like tower of massive states. Another possible generalization would be to consider a cubic array of M2-branes directly, rather than first reducing to string theory  along linear periodic array and then using the standard T-duality transformations on the remaining two directions. One might consider trying to perform a similar analysis with the BLG model \cite{Bagger:2006sk,Bagger:2007jr,Gustavsson:2007vu,Bagger:2007vi}. Since we are looking at an infinite number of M2-branes one need not restrict to a finite-dimensional 3-algebra. In particular one could consider an `affine' version of the so-called $\mathcal{A}_4$ 3-algebra:
\be
[T^a_m,T^b_n,T^c_p] = \lambda \varepsilon^{abcd}\delta_{de}T^e_{m+n+p}\ ,
\ee
where $a,b,c,d=1,2,3,4$,  $m,n,p\in {\mathbb Z}$ and $\lambda$ is arbitrary. It is easy to see that this satisfies all the conditions of a Euclidean Lie-3-algebra (totally antisymmetric, fundamental identity, positive definite metric). One might try to identify the resulting theory as describing an infinite array with two M2-branes located at each site. Another infinite dimensional 3-algebra is given by the Nambu 3-bracket:
\begin{equation}
\{  X , Y , Z  \}=  \epsilon^{i_1 i_2 i_3} \partial_{i_1}X \partial_{i_2}Y \partial_{i_3}Z\ .
\end{equation}
It has been suggested that a condensate of an infinite number of M2-branes describes an M5-brane through the Nambu 3-bracket \cite{Park:2008qe,Ho:2008nn}. The application considered here would lead to an interpretation of the Nambu 3-bracket in terms of an infinite array of M2-branes. However in the BLG model there are spurious massless fields in the Coloumb branch whenever the M2-branes are collinear \cite{Lambert:2008et}, as is the case here, so all modes appear massless. Nevertheless perhaps a suitable analysis could be made.

\section*{Acknowledgements}

The authors would like to thank the Isaac Newton Institute, Cambridge, where part of this work was completed.  The work by IJ is partially supported by NRF through the Korea-CERN theory collaboration and through the Center for Quantum Spacetime (CQUeST) of Sogang University with grant number 2005-0049409. NL would also like to thank the ICTS, Bangalore where this work was completed. PR is supported by the STFC studentship grant ST/F007698/1.

\bibliography{M2array_final_v2}

\providecommand{\href}[2]{#2}\begingroup\raggedright\begin{thebibliography}{10}

\bibitem{Bagger:2006sk}
J.~Bagger and N.~Lambert, ``{Modeling Multiple M2's},''
  \href{http://dx.doi.org/10.1103/PhysRevD.75.045020}{{\em Phys.Rev.}
  {\bfseries D75} (2007) 045020},
\href{http://arxiv.org/abs/hep-th/0611108}{{\ttfamily arXiv:hep-th/0611108
  [hep-th]}}.

\bibitem{Bagger:2007jr}
J.~Bagger and N.~Lambert, ``{Gauge symmetry and supersymmetry of multiple
  M2-branes},'' \href{http://dx.doi.org/10.1103/PhysRevD.77.065008}{{\em
  Phys.Rev.} {\bfseries D77} (2008) 065008},
\href{http://arxiv.org/abs/0711.0955}{{\ttfamily arXiv:0711.0955 [hep-th]}}.

\bibitem{Gustavsson:2007vu}
A.~Gustavsson, ``{Algebraic structures on parallel M2-branes},''
  \href{http://dx.doi.org/10.1016/j.nuclphysb.2008.11.014}{{\em Nucl.Phys.}
  {\bfseries B811} (2009) 66--76},
\href{http://arxiv.org/abs/0709.1260}{{\ttfamily arXiv:0709.1260 [hep-th]}}.

\bibitem{Bagger:2007vi}
J.~Bagger and N.~Lambert, ``{Comments on multiple M2-branes},''
  \href{http://dx.doi.org/10.1088/1126-6708/2008/02/105}{{\em JHEP} {\bfseries
  0802} (2008) 105},
\href{http://arxiv.org/abs/0712.3738}{{\ttfamily arXiv:0712.3738 [hep-th]}}.

\bibitem{Aharony:2008ug}
O.~Aharony, O.~Bergman, D.~L. Jafferis, and J.~Maldacena, ``{N=6 superconformal
  Chern-Simons-matter theories, M2-branes and their gravity duals},''
  \href{http://dx.doi.org/10.1088/1126-6708/2008/10/091}{{\em JHEP} {\bfseries
  0810} (2008) 091},
\href{http://arxiv.org/abs/0806.1218}{{\ttfamily arXiv:0806.1218 [hep-th]}}.

\bibitem{Bagger:2012jb}
J.~Bagger, N.~Lambert, S.~Mukhi, and C.~Papageorgakis, ``{Multiple Membranes in
  M-theory},''
\href{http://arxiv.org/abs/1203.3546}{{\ttfamily arXiv:1203.3546 [hep-th]}}.

\bibitem{Mukhi:2008ux}
S.~Mukhi and C.~Papageorgakis, ``{M2 to D2},''
  \href{http://dx.doi.org/10.1088/1126-6708/2008/05/085}{{\em JHEP} {\bfseries
  0805} (2008) 085},
\href{http://arxiv.org/abs/0803.3218}{{\ttfamily arXiv:0803.3218 [hep-th]}}.

\bibitem{Taylor:1996ik}
W.~Taylor, ``{D-brane field theory on compact spaces},''
  \href{http://dx.doi.org/10.1016/S0370-2693(97)00033-6}{{\em Phys.Lett.}
  {\bfseries B394} (1997) 283--287},
\href{http://arxiv.org/abs/hep-th/9611042}{{\ttfamily arXiv:hep-th/9611042
  [hep-th]}}.

\bibitem{Bashkirov:2010kz}
D.~Bashkirov and A.~Kapustin, ``{Supersymmetry enhancement by monopole
  operators},'' \href{http://dx.doi.org/10.1007/JHEP05(2011)015}{{\em JHEP}
  {\bfseries 1105} (2011) 015},
\href{http://arxiv.org/abs/1007.4861}{{\ttfamily arXiv:1007.4861 [hep-th]}}.

\bibitem{Martinec:2011nb}
E.~Martinec and J.~McOrist, ``{Monopole--Instantons in M2-Brane Theories},''
\href{http://arxiv.org/abs/1112.4073}{{\ttfamily arXiv:1112.4073 [hep-th]}}.

\bibitem{Ho:2009nk}
P.-M. Ho, Y.~Matsuo, and S.~Shiba, ``{Lorentzian Lie (3-)algebra and toroidal
  compactification of M/string theory},''
  \href{http://dx.doi.org/10.1088/1126-6708/2009/03/045}{{\em JHEP} {\bfseries
  0903} (2009) 045},
\href{http://arxiv.org/abs/0901.2003}{{\ttfamily arXiv:0901.2003 [hep-th]}}.

\bibitem{Kobo:2009gz}
T.~Kobo, Y.~Matsuo, and S.~Shiba, ``{Aspects of U-duality in BLG models with
  Lorentzian metric 3-algebras},''
  \href{http://dx.doi.org/10.1088/1126-6708/2009/06/053}{{\em JHEP} {\bfseries
  0906} (2009) 053},
\href{http://arxiv.org/abs/0905.1445}{{\ttfamily arXiv:0905.1445 [hep-th]}}.

\bibitem{Honma:2009bx}
Y.~Honma and S.~Zhang, ``{Quiver Chern-Simons Theories, D3-branes and
  Lorentzian Lie 3-algebras},''
  \href{http://dx.doi.org/10.1143/PTP.123.449}{{\em Prog.Theor.Phys.}
  {\bfseries 123} (2010) 449--474},
\href{http://arxiv.org/abs/0912.1613}{{\ttfamily arXiv:0912.1613 [hep-th]}}.

\bibitem{Nastase:2010ft}
H.~Nastase and C.~Papageorgakis, ``{Dimensional reduction of the ABJM model},''
  \href{http://dx.doi.org/10.1007/JHEP03(2011)094}{{\em JHEP} {\bfseries 1103}
  (2011) 094},
\href{http://arxiv.org/abs/1010.3808}{{\ttfamily arXiv:1010.3808 [hep-th]}}.

\bibitem{Douglas:2010iu}
M.~R. Douglas, ``{On D=5 super Yang-Mills theory and (2,0) theory},''
  \href{http://dx.doi.org/10.1007/JHEP02(2011)011}{{\em JHEP} {\bfseries 1102}
  (2011) 011},
\href{http://arxiv.org/abs/1012.2880}{{\ttfamily arXiv:1012.2880 [hep-th]}}.

\bibitem{Lambert:2010iw}
N.~Lambert, C.~Papageorgakis, and M.~Schmidt-Sommerfeld, ``{M5-Branes,
  D4-Branes and Quantum 5D super-Yang-Mills},''
  \href{http://dx.doi.org/10.1007/JHEP01(2011)083}{{\em JHEP} {\bfseries 1101}
  (2011) 083},
\href{http://arxiv.org/abs/1012.2882}{{\ttfamily arXiv:1012.2882 [hep-th]}}.

\bibitem{Park:2008qe}
J.-H. Park and C.~Sochichiu, ``{Taking off the square root of Nambu-Goto action
  and obtaining Filippov-Lie algebra gauge theory action},''
  \href{http://dx.doi.org/10.1140/epjc/s10052-009-1132-x}{{\em Eur.Phys.J.}
  {\bfseries C64} (2009) 161--166},
\href{http://arxiv.org/abs/0806.0335}{{\ttfamily arXiv:0806.0335 [hep-th]}}.

\bibitem{Ho:2008nn}
P.-M. Ho and Y.~Matsuo, ``{M5 from M2},''
  \href{http://dx.doi.org/10.1088/1126-6708/2008/06/105}{{\em JHEP} {\bfseries
  0806} (2008) 105},
\href{http://arxiv.org/abs/0804.3629}{{\ttfamily arXiv:0804.3629 [hep-th]}}.

\bibitem{Lambert:2008et}
N.~Lambert and D.~Tong, ``{Membranes on an Orbifold},''
  \href{http://dx.doi.org/10.1103/PhysRevLett.101.041602}{{\em Phys.Rev.Lett.}
  {\bfseries 101} (2008) 041602},
\href{http://arxiv.org/abs/0804.1114}{{\ttfamily arXiv:0804.1114 [hep-th]}}.

\end{thebibliography}\endgroup

\end{document}